\documentclass{PoS}
\pdfoutput=1 % if your are submitting a pdflatex (i.e. if you have
\usepackage{amsmath,amssymb,latexsym,bm,braket}
\usepackage{cite}
\usepackage{graphicx}
\usepackage{hepunits}
\usepackage{color}
\usepackage{multirow}

\allowdisplaybreaks

\title{QCD resummations for boosted top production}

\ShortTitle{QCD resummations for boosted top production}

\author{Andrea Ferroglia \\
       New York City College of Technology, 300 Jay Street\\
      Brooklyn, NY 11201, USA\\The Graduate School and University Center, The City University of New York\\ New York,
      NY 10016 USA\\
        E-mail: \email{aferroglia@citytech.cuny.edu}}

\author{\speaker{Benjamin D. Pecjak}\\
        Institute for Particle Physics Phenomenology,
      University of Durham\\
      DH1 3LE Durham, UK \\
        E-mail: \email{ben.pecjak@durham.ac.uk}}

\author{Darren J. Scott\\
        Institute for Particle Physics Phenomenology,
      University of Durham\\
      DH1 3LE Durham, UK \\
        E-mail: \email{d.j.scott@durham.ac.uk}}

\author{Li Lin Yang \\
     School of Physics and State Key Laboratory of Nuclear Physics and Technology\\
      Peking University, Beijing 100871, China\\
        E-mail: \email{yanglilin@pku.edu.cn}}

      \abstract{We present new results for QCD corrections to the
        top-pair invariant mass and top-quark $p_T$ distributions in
        boosted top-quark pair production at hadron colliders.  They
        are derived from a formalism which allows the joint
        resummation of soft and small-mass logarithms at NNLL$'$
        order, thus taking into account all potentially large
        corrections in the boosted regime, where the partonic
        center-of-mass energy is parameterically much larger than the
        mass of the top quark.  We match these results with those from
        standard soft-gluon resummation away from the small-mass limit
        to NNLL order and also with NLO fixed-order calculations, so
        that our results are valid in the maximum possible range of
        phase space. The resummation effects on the $p_T$ and top-pair
        invariant mass distributions are significant, bringing theory
        predictions into better agreement with experimental data compared 
        to pure NLO calculations.  }

\FullConference{8th International Workshop on Top Quark Physics, TOP2015\\
		14-18 September, 2015\\
		Ischia, Italy}

\begin{document}

\section{Introduction}

Detailed studies of top-quark properties are interesting for a number
of reasons (see \cite{delDuca:2015gca} for a recent review).  Many of
these are related to the large value of the top-quark mass, $m_t
\approx 173$~GeV.  For these
reasons, top-quark pair production is a benchmark process at hadron
colliders such as the LHC, and much work has gone into precision
Standard Model calculations to match the ever-growing accuracy of
experimental measurements.

Differential pair-production cross sections are largest in the region
of phase space characterized by $\hat{s}\gtrsim 4m_t^2$, with $\hat{s}$
the partonic center-of-mass energy squared.  In such regions of phase
space QCD corrections can be calculated straightforwardly as a
fixed-order expansion in the strong coupling constant $\alpha_s$.
Such fixed-order calculations have now been carried out to high
precision: results at next-to-next-to-leading order (NNLO) for
differential top-pair production cross sections were presented at this
conference and published recently in \cite{Czakon:2015owf}.  This
tremendously impressive calculation will form the baseline description
of top-pair production cross sections in perturbative QCD for years to
come, and adds to previous results for the total cross section
\cite{Czakon:2013goa} and the forward-backward asymmetry
at the Tevatron \cite{Czakon:2014xsa}.

While measurements of the total cross section and differential
distributions near their peaks are interesting, the collider
center-of-mass energies at LHC Run-I (7,8~TeV) and Run-II (13~TeV) are
sufficient to explore a qualitatively different region of phase space,
where the top quarks are produced with energies much larger than
their mass.  From a theoretical perspective, such ``boosted'' top-quark 
production is quantified by the condition that top-pairs are produced
in partonic collisions where $\hat{s} \gg m_t^2$. This
regime is especially interesting phenomenologically, as it is sensitive to the
%large energy scales needed to gain sensitivity to potentially heavy particles
%beyond the Standard Model.
possible new physics scale beyond the electroweak scale of the
Standard Model.  However, application of fixed-order perturbation
theory in the boosted regime is problematic due to the appearance of
potentially large corrections from soft and small-mass logarithms.

The purpose of this talk is to present a
resummation formalism tailored for QCD calculations of differential cross
sections in the boosted regime, and to briefly explore its implications on
phenomenology. The results presented are taken from a
detailed phenomenological analysis which is in progress
\cite{long_paper}.  The underlying theoretical basis for these results
was set up in \cite{Ferroglia:2012ku, Ferroglia:2013awa}, and some
higher-order perturbative results allowing precision applications were
derived in \cite{Ferroglia:2012uy, Broggio:2014hoa}.

\section{Mellin-space resummation for (boosted) top production}

In this section we present the resummation formalisms for both the
soft limit and the boosted soft limit.  QCD factorization allows one
to write the differential cross section with respect to the top-pair
invariant mass $M$ and the scattering angle as
\begin{equation}
	\frac{d^2\sigma(\tau)}{dM\,d\cos\theta}=\frac{8\pi\beta_t}{3sM}\sum\limits_{ij}\int^1_\tau \frac{dz}{z} \, \mathcal{L}_{ij}(\tau/z,\mu_f) \, C_{ij}(z,M,m_t,\cos\theta,\mu_f) \, ,
\label{eq:x-sec}
\end{equation}
where $z=M^2/\hat{s}$, $\tau=M^2/s$, $\beta_t=\sqrt{1-4m_t^2/M^2}$,
and $s$ is the hadronic center-of-mass energy squared. The
perturbatively calculable hard-scattering kernels $C_{ij}$ are related
to the partonic cross sections, with $ij$ denoting the partons in the
initial state. The parton luminosities $\mathcal{L}_{ij}$ are
non-perturbative functions defined as the convolution of parton
distribution functions (PDFs).
%\begin{equation}
%\mathcal{L}_{ij}(y,\mu_f) = \int_y^1 \frac{dx}{x} \, \phi_{i/N_1}(x,\mu_f) \, \phi_{j/N_2}(y/x,\mu_f) \, ,
%\end{equation}
%where $\phi_{i/N}$ is the parton distribution function (PDF) of the hadron $N$.

We shall consider the following two kinematic limits of the
hard-scattering kernels:
\begin{align}
\label{eq:softlimit}
\text{soft limit:} & \quad \hat{s},t_1,m_t^2 \gg \hat{s}(1-z)^2 \, ,
\\
\label{eq:boostedlimit}
\text{boosted soft limit:} & \quad \hat{s},t_1 
\gg m_t^2 \gg \hat{s}(1-z)^2  \gg m_t^2(1-z)^2 \,
\end{align}
In each of these cases the perturbative expansions of the
hard-scattering kernels contain large logarithms of scale ratios at
each order in perturbation theory. In particular, the presence of a
soft scale $\sqrt{\hat{s}}(1-z)$ introduces into the hard-scattering
kernel singular plus distributions of the form
\begin{align}
P_n(z) = \left[\frac{\ln^n (1-z)}{1-z} \right]_+ \, ,
\end{align}
and the presence of a collinear scale $m_t$ in the boosted limit leads
to large logarithms of the form $\ln^n(m_t^2/\hat{s})$. In the next
two subsections, we discuss formalisms to resum these logarithms to
all orders in perturbation theory.

For the discussion of resummation that follows, it is convenient to 
study the cross section in Mellin space.
We define the Mellin transform and its inverse by
\begin{align}
\tilde{f}(N) = \mathcal{M}[f](N) = \int_0^1 dx \,x^{N-1}{f}(x) \, , \quad f(x) = \mathcal{M}^{-1}[\tilde{f}](x) = \frac{1}{2 \pi i}\int_{c-i \infty}^{c+i \infty}dN \, x^{-N} \tilde{f}(N),
\label{mellint}
\end{align}
where the integration contour in the inverse transform is chosen such
that it lies to the right of all singularities in the function
$\tilde{f}(N)$. Convolutions such as the differential cross section in
Eq.~(\ref{eq:x-sec}) become simple products in Mellin space. Indeed,
the Mellin transform of Eq.~(\ref{eq:x-sec}) with respect to $\tau$
reads
\begin{align}
  \frac{d^2\widetilde{\sigma}(N)}{dM \, d\cos\theta}=
  \frac{8\pi\beta_t}{3sM}\sum_{ij}
  \widetilde{\mathcal{L}}_{ij}(N,\mu_f) \, \widetilde{c}_{ij}(N,M,m_t,\cos\theta,\mu_f)
  \, .
\label{eq:x-sec-mellin}
\end{align}
The limit $z \to 1$ corresponds to $N \to \infty$ in Mellin space,
with plus distributions related to logarithms of $\bar{N}\equiv N
e^{\gamma_E}$. In Mellin space, the soft and boosted soft limits in
Eq.~(\ref{eq:softlimit}) and (\ref{eq:boostedlimit}) are
\begin{align}
\label{eq:softlimit-mellin}
\text{Mellin-space  soft limit:} & \quad \hat{s},t_1,m_t^2 \gg \frac{\hat{s}}{N^2} \, ,
\\
\label{eq:boostedlimit-mellin}
\text{Mellin-space boosted soft limit:} & \quad \hat{s},t_1 
\gg m_t^2 \gg \frac{\hat{s}}{N^2} \gg \frac{m_t^2}{N^2} \,.
\end{align}

We finally note that while Eq.~(\ref{eq:x-sec}) is written for the
differential cross section with respect to the top-pair invariant mass
$M$, in the soft limit one can easily perform a change of variable to
obtain differential cross section with respect to the top quark
transverse momentum $p_T$. Of course contributions away from the soft
limit are different in the two cases, but we can take them into
account at fixed-order through a matching procedure to be discussed
later.

\subsection{The soft limit}

\noindent
In the soft limit, the resummed hard-scattering kernel can be written as
\begin{multline}
\label{eq:soft-resummed}
\widetilde{c}_{ij}(N,M,m_t,\cos\theta,\mu_f) = 
{\rm Tr} \Bigg[\widetilde{\bm{U}}^m_{ij}(\mu_f,\mu_h,\mu_s) \, \bm{H}^m_{ij}(M,m_t,\cos\theta,\mu_h) \, \widetilde{\bm{U}}_{ij}^{m\dagger}(\mu_f,\mu_h,\mu_s)
\\ 
\times \widetilde{\bm{s}}^m_{ij}\left(\ln\frac{M^2}{\bar{N}^2 \mu_s^2},M,m_t,\cos\theta,\mu_s \right)\Bigg] + \mathcal{O}\left(\frac{1}{N}\right) .
\end{multline}
The formula consists of two main elements. Firstly, the hard functions
$\bm{H}^m_{ij}$ and the soft functions $\widetilde{\bm{s}}^m_{ij}$ are
matching functions which contain contributions from the two widely
separated scales in (\ref{eq:softlimit-mellin}).  Secondly, the
evolution factors $\widetilde{\bm{U}}^m_{ij}$ arise from solving the
renormalization group (RG) equations of the hard and soft functions,
and can be written as path-ordered exponentials of the anomalous
dimensions governing their evolution.
For details about these functions, we refer the readers to \cite{Ahrens:2010zv}.

The philosophy of RG-improved perturbation theory is to choose the
hard and soft scales around their natural values $\mu_h\sim M$ and
$\mu_s\sim M/\bar{N}$, such that large logarithms are absent from
the hard and soft functions. These large logarithms are exponentiated
into the RG evolution factors, and the resummation of them is thus
achieved. The logarithmic orders are counted by treating the large
logarithms such as $\ln(\mu_h/\mu_s)$ as
$\mathcal{O}(1/\alpha_s)$. For next-to-next-to-leading logarithmic
(NNLL) accuracy, we need the hard and soft functions to 1-loop, the
various anomalous dimensions to 2-loop, and the 3-loop cusp anomalous
dimension, which were calculated or collected in
\cite{Ahrens:2010zv}. 
%Note however that \cite{Ahrens:2010zv} used a
%different choice for the soft scale $\mu_s$, as for momentum 
%space resummation. 

\subsection{The boosted soft limit}

The boosted soft limit (\ref{eq:boostedlimit}) can be regarded as the
$m_t \to 0$ limit of the soft limit. Taking $m_t \to 0$ in the
resummation formula (\ref{eq:soft-resummed}), one sees that both the
hard and soft functions develop collinear logarithms of the form
$\ln^n(m_t/M)$. It is possible to show that the hard and soft
functions can be further factorized in this limit as
\begin{align}
 \bm{H}^m_{ij}(M,m_t,\cos\theta,\mu_f) &= \bm{H}_{ij}(M,\cos\theta,\mu_f) \, C_D^2(m_t,\mu_f) + \mathcal{O} \left( \frac{m_t}{M} \right) ,
 \\
 \widetilde{\bm{s}}^m_{ij} \left( \ln\frac{M^2}{\bar{N}^2 \mu_f^2},M,m_t,\cos\theta,\mu_f \right) &= \widetilde{\bm{s}}_{ij} \left( \ln\frac{M^2}{\bar{N}^2 \mu_f^2},M,\cos\theta,\mu_f \right) \, \widetilde{s}_D^2 \left( \ln\frac{m_t}{\bar{N}\mu_f}, \mu_f \right) + \mathcal{O} \left( \frac{m_t}{M} \right) .
\end{align}
In the above formulas, the hard functions $\bm{H}_{ij}$ and soft
functions $\widetilde{\bm{s}}_{ij}$ without the superscript $m$ are
independent of the top quark mass $m_t$. They were calculated to NNLO
in \cite{Broggio:2014hoa} and \cite{Ferroglia:2012uy},
respectively. All the $m_t$-dependence is now factorized into the two
functions $C_D$ and $\widetilde{s}_D$, which are related to the
perturbative heavy-quark fragmentation function, and were extracted at
NNLO in \cite{Ferroglia:2012ku}.  Again solving RG equations for the
component functions, the result for the resummed hard scattering
kernel in the boosted soft limit is\footnote{In the presence of
  heavy-quark loops the factorization of the partonic cross sections
  is more involved, therefore, we add such contributions onto the
  resummation formula (\ref{eq:boosted-resummed}) using fixed-order
  perturbation theory.}
\begin{multline}
\label{eq:boosted-resummed}
\widetilde{c}_{ij}(N,M,m_t,\cos\theta,\mu_f) = 
{\rm Tr} \Bigg[\widetilde{\bm{U}}_{ij}(\mu_f,\mu_h,\mu_s) \, \bm{H}_{ij}(M,\cos\theta,\mu_h) \, \widetilde{\bm{U}}_{ij}^{\dagger}(\mu_f,\mu_h,\mu_s)
\\ 
\times \widetilde{\bm{s}}_{ij}\left(\ln\frac{M^2}{\bar{N}^2 \mu_s^2},M,\cos\theta,\mu_s \right)\Bigg] 
\, \widetilde{U}_D^2(\mu_f,\mu_{dh},\mu_{ds})
\, C_D^2(m_t,\mu_{dh})\, \widetilde{s}_D^2 \Biggl(
  \ln\frac{m_t}{\bar{N}\mu_{ds}}, \mu_{ds} \Biggr)
\\ 
+ \mathcal{O} \left( \frac{1}{N} \right) + \mathcal{O} \left( \frac{m_t}{M} \right) .
\end{multline}
The evolution factors $\widetilde{\bm{U}}_{ij}$ and $\widetilde{U}_D$
resum large logarithms when the matching scales are chosen to be near
their canonical values: $\mu_h \sim M$, $\mu_s \sim M/\bar{N}$,
$\mu_{dh}\sim m_t$, $\mu_{ds}\sim m_t/\bar{N}$. This resummation can
be performed at NNLL accuracy as in the soft case discussed
earlier. However, since the component functions in the boosted soft
limit are known to NNLO in perturbation theory, we can also
incorporate them into the resummation formula. We follow common
nomenclature and refer to this accuracy as NNLL$'$, where the $'$
emphasizes that the matching functions are known to one order higher
than needed in a pure NNLL calculation (such as that in the soft limit
described above).

\section{Matching across kinematic limits}

The resummations derived above are based on factorization formulas
applicable in the soft or boosted soft limit.  These factorization
formulas are only valid up to leading order in their relevant
expansion parameters, so every time a resummation is performed some
information on subleading corrections is thrown away.  However,
these subleading corrections can be taken into account through a
matching procedure.  In our case, the optimal combination of the
resummed formulas in the boosted soft and soft limit can be combined
with fixed-order calculations according to
\begin{align}
\label{eq:fully-matched}
d \sigma^{\mathrm{NLO+NNLL'}} = d\sigma^{\mathrm{NNLL'}^b} 
& + \left(d\sigma^{\mathrm{NNLL}^m} - 
\left. d\sigma^{\mathrm{NNLL}^b}\right|_{\substack{\mu_{\mathrm{ds}}=\mu_s \\
			\mu_{\mathrm{dh}=\mu_h}}} \right)
			%\nonumber \\ & 
+ \left(d\sigma^{\mathrm{NLO}} - 
\left. d\sigma^{\mathrm{NNLL}^m}\right|_{\substack{\mu_{\mathrm{s}}=\mu_f \\
			\mu_{\mathrm{h}=\mu_f}}} \right) 		\,.	 
\end{align}
The logarithmic counting has been defined above, and the superscripts
$b$ and $m$ refer to results valid in the boosted soft and soft limits
respectively.  The first term in the above equation is the NNLL$'$
resummation formula in the boosted soft limit. The difference of terms
in the first bracket takes into account corrections from NNLL
resummation in the soft limit which vanish in the limit $m_t\to 0$ and
are therefore not contained in the first term. Finally, the difference
of terms in the second bracket takes into account corrections from
fixed-order perturbation theory which are subleading in the $z\to 1$
limit and thus accounted for by neither of the first two terms.  It is
straightforward to extend the above procedure to incorporate the
recent NNLO calculations in \cite{Czakon:2015owf}; however, we prefer
to do this only once predictions with the dynamical scale choices used in
our numerical calculations in the next section (i.e. not $\mu_f=m_t$)
are available.

\begin{figure}[t!]
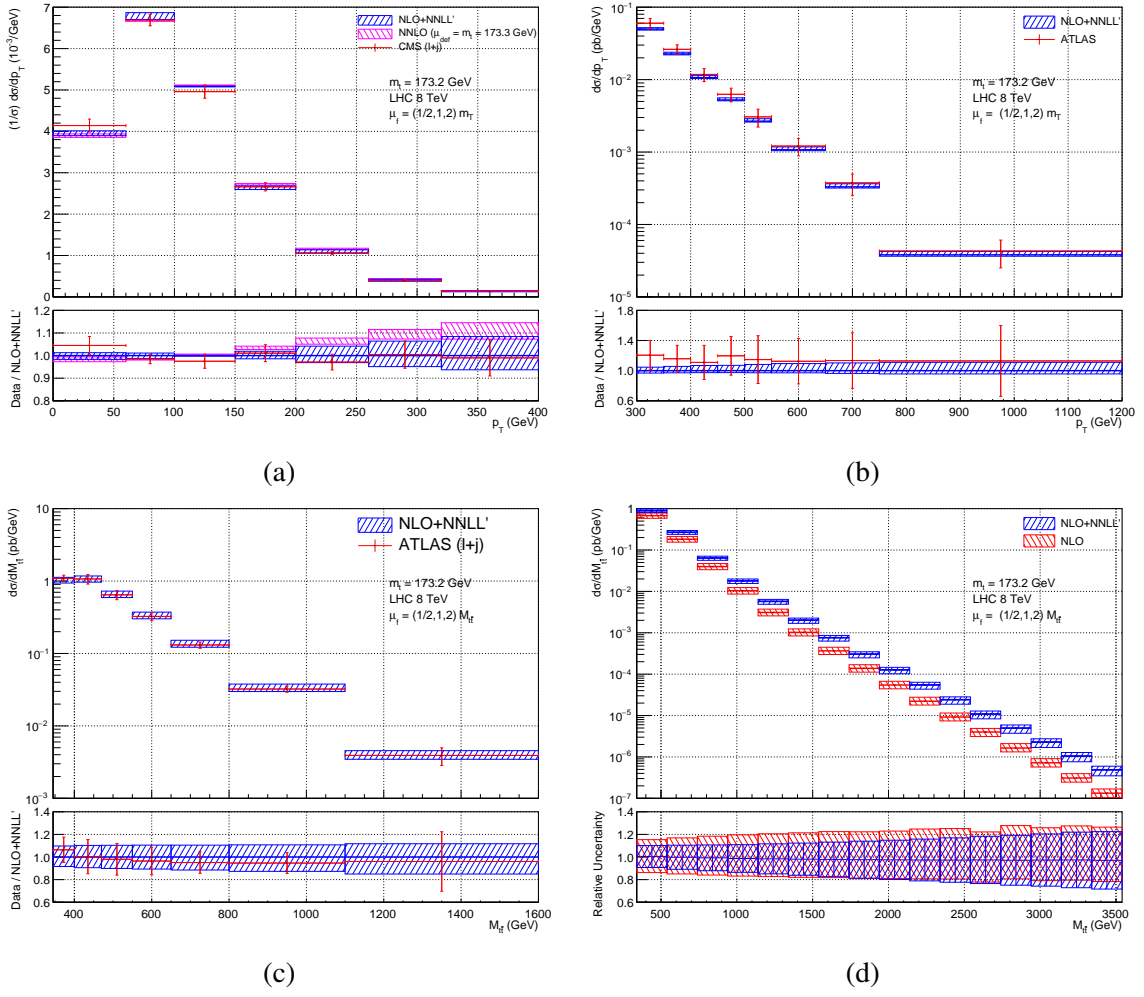

\centering
\begin{tabular}{cc}
\includegraphics[width=0.48\textwidth]{{{pTt_lhc8_cms_0.50}}}
&
\includegraphics[width=0.48\textwidth]{{{pTt_lhc8_atlas_boosted_0.50}}}
\\
(a) & (b)
\\
\includegraphics[width=0.48\textwidth]{{{Mtt_lhc8_atlas_1.00}}}
&
\includegraphics[width=0.48\textwidth]{{{Mtt_lhc8_1.00}}}
\\
(c) & (d)
\end{tabular}
\caption{Numerical results as explained in the text.  Note that while 
the bottom panels of (a)-(c) display ratios of predictions, that in (d)
displays relative uncertainties.\label{fig:results}}
\end{figure}

\section{Phenomenology}

We now provide numerical results derived from the resummation formulas
described above.  These results are obtained by first evaluating the
Mellin transformed cross section (\ref{eq:x-sec-mellin}) at a given
point in phase space, and then performing the inverse Mellin transform
numerically using the Minimal Prescription \cite{Catani:1996yz}.  This
procedure requires an efficient construction of Mellin-transformed
parton luminosities: for this we use methods described in
\cite{Bonvini:2014joa, Bonvini:2012sh}.

In all our numerics we choose $m_t=173.2$~GeV and use MSTW2008NNLO
PDFs \cite{Martin:2009iq}. The default matching scales are fixed at
their canonical values listed above.  The factorization scale is
correlated with the observables under consideration: for the invariant
mass distribution its default value is $\mu_f=M$, while for the
transverse momentum distribution it is set equal to the transverse
mass $\mu_f=m_T\equiv \sqrt{p_T^2+m_t^2}$. The scale uncertainties are
estimated by varying all scales independently by factors of two around their default values and
combining the resulting variations in quadrature.
The results of our analysis are shown in Figure~\ref{fig:results}.

In Figure~\ref{fig:results} (a), we show our NLO+NNLL$'$ prediction for the
top-quark transverse momentum distribution in the relatively low
energy (unboosted) region, compared  with a recent experimental
measurement in the lepton+jet channel by the CMS collaboration
\cite{Khachatryan:2015oqa} and also the NNLO results from
\cite{Czakon:2015owf}.\footnote{We note that \cite{Czakon:2015owf} uses
fixed scales with default values $\mu_r=\mu_f=m_t$, and has chosen a
slightly different top quark mass $m_t=173.3$~GeV.} One 
sees that the NLO+NNLL$'$ prediction agrees very well with the data, producing
a slightly softer distribution at higher-$p_T$ than the NNLO calculation.

In Figure~\ref{fig:results} (b), we present the top-quark transverse
momentum distribution in the high-$p_T$ (boosted) region, where both
the soft and the small-mass logarithms resummed through our formalism are
expected to be relevant. We compare our results with a recent
experimental measurement by the ATLAS collaboration \cite{Aad:2015hna}
carried out by using fat-jet techniques especially designed for 
boosted kinematics. We find that our prediction agrees well with the
data, although the experimental uncertainties are rather large.

In Figure~\ref{fig:results} (c), we compare the top-pair invariant
mass distribution with measurements from the ATLAS collaboration
\cite{Aad:2015hna}.  The two are in good agreement, although
especially in the highest invariant-mass bin the experimental
uncertainties are  significant.  

Finally, in Figure~\ref{fig:results} (d), we display results reaching
up to very high top-pair invariant mass.  The results show significant
resummation effects compared to NLO (where $\mu_f=M$ by default),
although as seen in the bottom panel of that plot the relative
uncertainties are roughly equal.

Run-II of the LHC will produce much more data in the boosted regime,
which will allow for higher-precision comparisons of predictions and data
for differential distributions.  It will also be interesting to
compare (and above all, match) with the NNLO results once they are
available in the high-$p_T$ region with dynamical scale settings.

\end{document}